\documentclass[aps,prl,twocolumn,showpacs]{revtex4} 
\usepackage{amsmath} 
\usepackage{graphicx} 
\usepackage{color}
\usepackage{verbatim}
\usepackage{bm}
\usepackage{epstopdf}
\usepackage[bookmarks=false,colorlinks=true,urlcolor=blue,citecolor=blue,linkcolor=blue]{hyperref}

\renewcommand{\vec}[1]{{\bm #1}}

\newcommand{\addMR}[1]{\textcolor{black}{#1}} 

\begin{document} 
\title{Composite topological excitations in ferromagnet-superconductor heterostructures}

\author{Kjetil M. D. Hals, Michael Schecter and Mark S. Rudner} 
\affiliation{Niels Bohr International Academy and the Center for Quantum Devices, Niels Bohr Institute, University of Copenhagen, 2100 Copenhagen, Denmark}
%%%%%%%%%%%%%%%%%%%%%%%%%%%%%%%%%%%%%%%%%%%%%%%%%%%%%%%%%%%%%%%%%%%%%%%%%%%%%%% 
\begin{abstract}
We investigate the formation of a new type of composite topological excitation -- the skyrmion-vortex pair (SVP) -- in hybrid systems consisting of coupled ferromagnetic and superconducting layers.
Spin-orbit interaction in the superconductor mediates a magnetoelectric coupling between the vortex and the skyrmion, with a sign (attractive or repulsive) 
that depends on the topological indices of the constituents. We determine the conditions under which a bound SVP is formed, and characterize the range and depth of the effective binding potential through analytical estimates and numerical simulations. Furthermore, we develop a semiclassical description of the coupled skyrmion-vortex dynamics and discuss how SVPs can be controlled by 
applied spin currents. 
\end{abstract}

\maketitle 

%%%%%%%%%%%%%%%%%%%%%%%%%%%%%%%%%%%%%%%%%%%%%%%%%%%%%%%%%%%%%%%%%%%%%%%%%%%%%%% 
%\section{Introduction} 
%%%%%%%%%%%%%%%%%%%%%%%%%%%%%%%%%%%%%%%%%%%%%%%%%%%%%%%%%%%%%%%%%%%%%%%%%%%%%%% 
Advances in materials and fabrication capabilities in recent years have opened many possibilities for modifying and harnessing the properties of matter in a variety of interesting and powerful ways.
In particular, {\it hybrid systems} comprised of layers of two or more materials of very different character provide new opportunities to study important fundamental phenomena -- such as magnetism and superconductivity~\cite{Reviews:MajoranaWires,Reviews:SuperSpintronics, Dagan:LAOSTO, Moler:LAOSTO}, or optical and electronic properties~\cite{KoppensPhotodetectorReview} -- in new regimes and in new combinations of coexistence.

Often, the range of phenomena exhibited by a hybrid system is much richer than that of its parts~\cite{GeimReview}.
For example, it was recently demonstrated that hybrid systems comprised of superconductors and semiconductors yield exquisite new levels of fully-electrical control over Josephson-based quantum devices~\cite{gatemon}.
It has also been proposed that the exchange field of a magnetic layer proximity-coupled to a 3D topological insulator surface may open the 
possibility to realize a new quantum phase of matter with an intriguing quantized magneto-electric response \cite{QiHugheZhong2008}.
Moreover, the possibilities afforded by the trifold combination of magnetic, superconducting, and semiconducting systems is at the heart of the intense wave of recent activity aimed at realizing topological superconductivity and associated Majorana bound states~\cite{Kitaev:2001, MajoranaReviews,  Sato:prb2010, Sau:prl2010, Nakosai:prb2013, Braunecker2013, Klinovaja2013, Vazifeh2013} -- a key step in the development of topological quantum information processing~\cite{Nayak:RMP2008}.

%%%%%%%%%%
\begin{figure}[t!] 
\centering 
\includegraphics[scale=1.0]{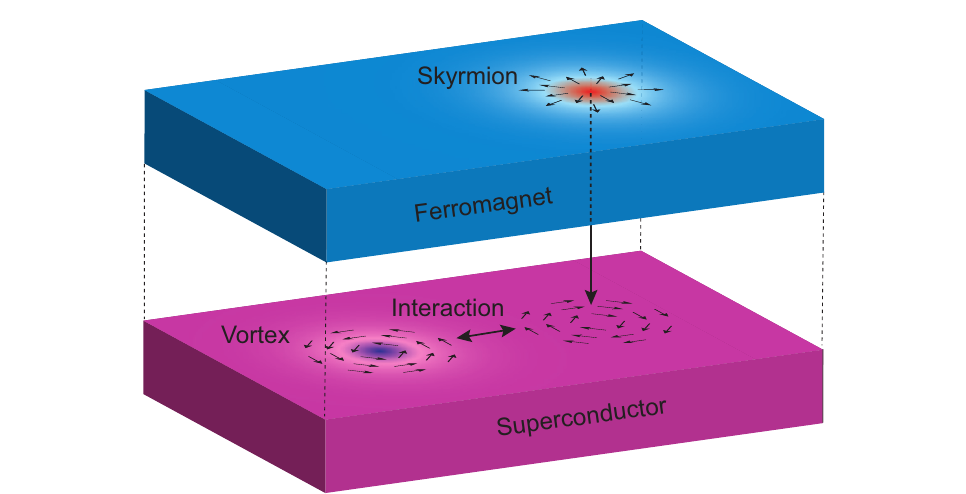}  
\caption{(Color online). The exchange field of a ferromagnetic thin film is induced 
into a thin film superconductor by proximity coupling. The spatially varying exchange field of a skyrmion (radially oriented arrows in the ferromagnet), generates a supercurrent in the superconductor (circulating arrows) due to the magnetoelectric effect.  The interaction of magnetoelectric and vortex-induced currents leads to the binding of skyrmion-vortex pairs (SVPs). }
\label{Fig1}\end{figure}
%%%%%%%%%%

In this work, we investigate a novel type of composite topological excitation in hybrid systems.
Specifically, we investigate the coupling between magnetic skyrmions and superconducting vortices in a two-dimensional (2D) layered ferromagnet-superconductor heterostructure (see Fig.~\ref{Fig1}).
The combination of spin-orbit coupling (SOC) in the superconductor and the lack of inversion symmetry of the heterostructure leads to a magnetoelectric coupling that mediates an interaction between textures of the magnetic and superconducting order parameters. In isolation, both the superconductor and the 2D ferromagnet may host a variety of robust topological excitations -- vortices and anti-vortices for the superconductor \cite{Blatter94} and skyrmions of variable helicity and chirality for the ferromagnet~\cite{SkyrmionReviews}. When these systems are brought together in a heterostructure, we find that certain combinations of these entities bind to form a new type of composite topological excitation, which we refer to as the skyrmion-vortex pair (SVP). 

After establishing the stability of SVPs and describing the conditions under which they form, we develop a semiclassical theory to describe the dynamics of the pair.
Motion of the composite SVP can be driven, e.g., by externally applied spin torques caused by electric currents~\cite{SkyrmionReviews, Kajiwara:nature10} or thermally excited spin waves~\cite{Kong:prl13},  which act on the skyrmionic component of the SVP. We derive conditions on the strength of the external forcing which ensure that the SVP remains bound and moves as a unit.
The motion can for example be detected through direct imaging via nanoscale scanning magnetometry~\cite{Zeldov:NanoSQUID, Thiel:arxiv15}. 

\emph{Coupling mechanism} $\--$ 
We expect SVPs to form in materials with strong SOC and broken spatial inversion symmetry.  Here, the skyrmion-vortex interaction is primarily mediated through the superconducting magnetoelectric effect~\cite{edelstein1,edelstein2,Yip, balatsky}, whereby an induced spin polarization generates a supercurrent.  In the case of ferromagnet-superconductor heterostructures, in which inversion symmetry is broken by the interface, the linear coupling between the magnetic exchange field 
$\mathbf{h}({\bf r})= h_0 \hat{\mathbf{h}} ({\bf r})$ and the supercurrent $\sim \boldsymbol{\nabla} \theta_{\rm s}$ ($\theta_{\rm s}$ being the superconducting phase) is modeled by the Lifshitz invariant \cite{edelstein2, balatsky}
\begin{equation}
F_{\rm me}= \kappa\int {\rm d}\mathbf{r}\,  \left(\hat{\mathbf{z}}\times\mathbf{h} \right)\cdot \left(\boldsymbol{\nabla}\theta_{\rm s}/2+\bf{A}\right),
\label{Eq:Fme}
\end{equation}
in the free energy. Here, ${\bf A}$ is the magnetic vector potential (we set $e=\hbar=c = 1$). The parameter $\kappa$ is proportional to the SOC and $\hat{\bf z}$ is perpendicular to the interface. The skyrmion and vortex produce spatially varying exchange and phase fields, which through Eq.~\eqref{Eq:Fme} give rise to their mutual interaction.
 
The form of the interaction can be determined by recalling that a vortex in the superconductor induces a winding phase field:  $\boldsymbol{\nabla}\theta_{\rm s}=q_v\hat{\vec\phi}_v/r_v$, where $(r_v,\,\phi_v)$ are polar coordinates in the frame where the vortex core lies at the origin, and $q_v$ is the vorticity. The profile of a skyrmion at the origin
can be written in the 
form
\begin{align}
\label{eq:skyrmion}
\hat{\mathbf{h}}=\left(\cos\Phi(\mathbf{r})\sin\Theta(\mathbf{r}),\,\sin\Phi(\mathbf{r})\sin\Theta(\mathbf{r}),\cos\Theta(\mathbf{r})\right),
\end{align}
with the boundary conditions $\Theta(0)=\pi,\,\Theta(\infty)=0$.
The skyrmion is characterized by a topological index $q_s=\frac{1}{4\pi}\int{\rm d}{\bf r}\,\hat{{\bf h}}\cdot(\partial_x\hat{{\bf h}}\times\partial_y\hat{{\bf h}})$; for the profile above, 
$q_s$ is just the winding number of $\Phi(\mathbf{r})$ along a loop enclosing the skyrmion core \cite{SkyrmionReviews}. Below we assume for simplicity  $\Phi(r,\phi)=q_s\phi+\varphi$, where $\varphi$ is the skyrmion helicity \cite{SkyrmionReviews} and $(r, \phi)$ are polar coordinates in the frame where the origin lies at the skyrmion core. 

We focus on the case of a thin film type-II superconductor where the penetration depth $\lambda$ can greatly exceed the size of the vortex and skyrmion cores \cite{GubinPRB2005}. 
Here, the screening currents, ${\bf j}= -{\bf A}/4\pi\lambda^2$, induced by orbital or dipolar magnetic fields may be 
neglected. Therefore, below we set ${\bf A} = 0$. Substituting the above exchange and phase profiles into Eq.~\eqref{Eq:Fme} gives
\begin{equation}
\label{Eq:Fme2}
F_{\rm me} (r_{\rm sv}; q_s = 1) = \kappa q_vh_0 R_{\rm s} f(r_{\rm sv}) \cos \varphi,
\end{equation}
where $r_{\rm sv}$ is the skyrmion-vortex separation, $R_{\rm s}$ is the skyrmion core size, and the dimensionless function $f(r_{\rm sv})=\frac{\pi}{R_{\rm s}}\int_{r_{\rm sv}}^\infty dr \sin\Theta(r)$ depends on the precise shape of the skyrmion profile and is monotonic when $\sin\Theta(r)$ is sign definite.
Most essentially, $f(r_{\rm sv})$ approaches zero rapidly once $r_{\rm sv}\gtrsim R_{\rm s}$.

The skyrmion-vortex interaction in Eq.~\eqref{Eq:Fme2} can be attractive or repulsive, with the sign of $F_{\rm me}$ controlled by the vorticity $q_v$, the skyrmion helicity $\varphi$, and the overall sign of the exchange field, $h_0$.
Skyrmion-vortex binding is expected  when $\kappa q_vh_0 \cos \varphi<0$.  For a N\'{e}el (hedgehog) skyrmion, where $\varphi=0$ or $\pi$, the interaction depends on the sign of $\kappa q_v h_0$, while for a Bloch (spiral) skyrmion, where the in-plane field is rotated by $\pi/2$ ($\varphi=\pm\pi/2$), the interaction vanishes to linear order in Rashba SOC.

Physically, Eq.~\eqref{Eq:Fme2} can be understood in terms of the mutual current-current interaction between the skyrmion and vortex. While the vortex gives rise to a circulating current pattern, the skyrmionic exchange field induces the current ${\bf j}_{\rm me}=\kappa \left(\hat{z}\times {\bf h}\right)=\kappa h_0\sin\Theta(\cos\varphi\,\hat{\vec\phi}-\sin\varphi\,\hat{\vec r})$, due to the magnetoelectric effect \cite{Yip, balatsky}. As a result, the energy density ${\bf j}_{\rm me} \cdot \boldsymbol{\nabla}\theta_{\rm s}$ is lowered when the induced currents flow in opposite directions. 

In the case of the Bloch skyrmion $(\varphi=\pm\pi/2)$ there is a residual current-current interaction proportional to the {\em square} of the SOC. 
This term is given by~\cite{balatsky}: 
$F^{(2)}\propto\int {\rm d\mathbf{r}}\, (\hat{{\bf z}}\times \boldsymbol{\nabla}h_z) \cdot \boldsymbol{\nabla} \theta_{\rm s}/2 = -\pi q_v h_z(r_{\rm sv})$. 
The current produced by the skyrmion in this case, ${\bf j}^{(2)}\propto \hat{{\bf z}}\times \boldsymbol{\nabla}h_z=h_0 \partial_r\cos\Theta\,\hat{\vec{\phi}}$, leads to an attractive interaction for $q_v h_0 <0$.

%%%%%%%%%%%%%%%%%%%%%%%%%%%%%%%%%%%%%%%%%%%%%%%%%%%%%%%%%%%%%%%%%%
\begin{table}[t] 
\centering
\begin{tabular}{c| c| c }  
\hline\hline 
                                             & Vortex $q_v>0$ &    Antivortex $q_v<0$ \\ 
\hline 
\,\,\,N\'{e}el $\varphi=0$\,\,\,\, & Att.  &  Rep.              \\ 
\hline 
N\'{e}el $\varphi=\pi$              &        Rep.      &  Att.          \\
\hline 
Bloch        $\varphi=\pi/2$        & Att. & Rep.                  \\
\hline\hline
\end{tabular}
\caption{Summary of the skyrmion-vortex interaction.  Att.~and Rep. indicate attraction and repulsion, respectively. In all cases, we assume $\kappa >0$ and $h_0<0$ [see Eq.~\eqref{Eq:Fme2}]. } 
\label{table1}
\end{table}
%%%%%%%%%%%%%%%%%%%%%%%%%%%%%%%%%%%%%%%%%%%%%%%%%%%%%%%%%%%%%%%%%%

Throughout this work we focus on the case $q_s=1$, where  the skyrmion profile is invariant with respect to angular rotations about the skyrmion core \cite{SkyrmionReviews}. 
For $q_s\neq 1$ the exchange profile has periodic angular modulations that lead to a skyrmion-vortex interaction with a {\it sign that depends on direction from the skyrmion core (rather than just the separation $r_{\rm sv}$).} 
In this case, skyrmion-vortex binding is not expected for a system with Rashba SOC~\cite{Comment:Dresselhaus}.  Our predictions for skyrmion-vortex binding in this case are summarized in Table \ref{table1}.

%%%%%%%%%%%%%%%%%%%%%%%%%%%%%%%%%%%%%%%%%%%%%%%%%%%%%%%%%%%%%%%%%%%%%%%%%%%%%%%%%%%%%%%%%%%%%%%%%%%%%%%%%%%%%%%% 
%Model and method : 
%%%%%%%%%%%%%%%%%%%%%%%%%%%%%%%%%%%%%%%%%%%%%%%%%%%%%%%%%%%%%%%%%%%%%%%%%%%%%%%%%%%%%%%%%%%%%%%%%%%%%%%%%%%%%%% %
We now support the predictions above with  a microscopic model. 
We study the skyrmion-vortex interaction using the two-dimensional tight-binding Hamiltonian~\addMR{(see, e.g., Ref.~\cite{balatsky})}
\begin{eqnarray}
H &=& -t \sum_{\langle ij \rangle} \mathbf{c}_{i}^{\dagger} \mathbf{c}_{j} - \mu\sum_{i} \mathbf{c}_{i}^{\dagger} \mathbf{c}_{i} - \sum_{i}  \mathbf{c}_{i}^{\dagger} \left(  \mathbf{h}_i \cdot\boldsymbol{\sigma} \right) \mathbf{c}_{i} + \label{Eq:H0} \\
& &   i\alpha_R \sum_{\langle ij \rangle}  \mathbf{c}_{i}^{\dagger}\hat{\mathbf{z}}\cdot (  \hat{\mathbf{d}}_{ij} \times \boldsymbol{\sigma} ) \mathbf{c}_{j} + \sum_{i} \left( \Delta_i  c_{i\uparrow}^{\dagger} c_{i\downarrow}^{\dagger} +  h.c. \right)  .    \nonumber
\end{eqnarray}
Here, $\mathbf{c}^{\dagger}_i  = (c_{i \uparrow}^{\dagger}  \  c_{i \downarrow}^{\dagger} )$, where $c_{i \sigma}^{\dagger}$ creates an electron with spin $\sigma$ at lattice site $i= (x,y)$.
The symbol $\langle ij \rangle$ indicates a summation over nearest-neighbor lattice sites and $ \hat{\mathbf{d}}_{ij} $ is a unit vector that points from site $j$ to site $i$. 
The first term in Eq.~\eqref{Eq:H0} describes electron hopping between neighboring lattice sites with matrix element $t$, $\mu$ is the chemical potential, $\mathbf{h}$ is the exchange field induced by the ferromagnet, $\boldsymbol{\sigma}$ is the vector of Pauli matrices, and $\alpha_R$ parametrizes the Rashba SOC. The last term describes superconducting s-wave pairing, where the pair potential $\Delta_i$ is calculated self-consistently \cite{Sacramento:prb07, deGennes:book, supp}. 

The skyrmion is introduced via the discretized exchange field ${\bf h}_i={\bf h}({\bf r}_i)$, Eq.~\eqref{eq:skyrmion}, where $\vec{r}_i$ is the spatial coordinate of lattice site $i$ and for demonstration we take $\cos\Theta(\mathbf{r})=\frac{r^2- R_{\rm s}^2}{r^2+R_{\rm s}^2}$ with $R_{\rm s}=2a$ ($a$: the lattice constant). To introduce a vortex in the superconductor, we initialize the self-consistency calculation for the pairing potential using $\Delta_i= |\Delta|\exp (iq_v \phi_{v, i})$~\cite{Comment:h}. 

%%%%%%%%%%%%%%%%%%%%%%%%%%%%%%%%%%%%%%%%%%%%%%%%%%%%%%%%%%%%%%%%%%%%%%%%%%%%%%% 
%Results and discussion
%%%%%%%%%%%%%%%%%%%%%%%%%%%%%%%%%%%%%%%%%%%%%%%%%%%%%%%%%%%%%%%%%%%%%%%%%%%%%%% 

%%%%%%%%%%%%%%%%%%%%%%%%%%%%%%%%%%%%%%%%%%%%%%%%%%%%%%%%%%%%%%%%%%%%%%%%%%%%%%%%%%% 
\begin{figure}[t!] 
\centering 
\includegraphics[scale=1.0]{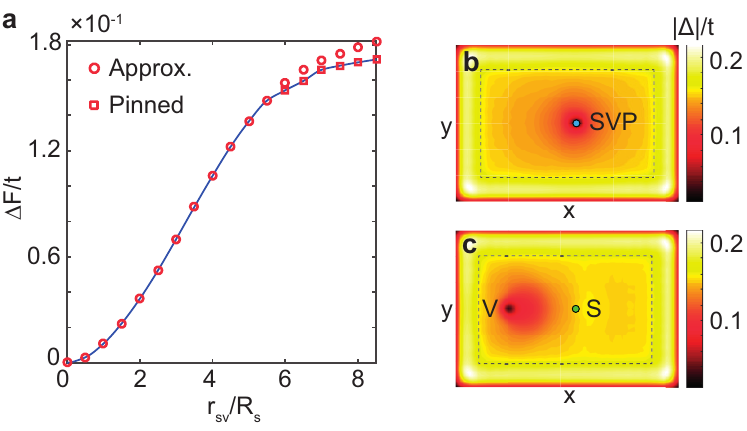}  
\caption{(Color online). {\bf (a)} Change of the free energy $\Delta F (r_{\rm sv})= F(r_{\rm sv}) - F_{\rm eqv}$ {\it vs.} skyrmion-vortex separation $r_{\rm sv}$, for a N\'{e}el skyrmion ($\varphi=0$ and $\kappa h_0 q_v<0$). $F_{\rm eqv}$ is the equilibrium value of the free energy at $r_{\rm sv}=0$.  The circles represent values calculated from Eq.~\eqref{Eq:F} using 
a fixed vortex profile obtained  self-consistently at $r_{\rm sv}=0$. The squares represent a pinned vortex whose shape is determined self-consistently for each separation. The line is a guide to the eye. 
 {\bf (b)-(c)} Ground state order parameter profiles in the presence of a N\'{e}el skyrmion with fixed location. {\bf (b)}  For $\varphi=0$, the skyrmion (S) and vortex (V) form a bound SVP, with their cores lying on top of each other. {\bf (c)} For  $\varphi=\pi$, the vortex is repelled from the skyrmion and is pushed to the edge of the ferromagnetic region.}
\label{Fig3} 
\end{figure} 
\begin{comment}
\end{comment}
%%%%%%%%%%%%%%%%%%%%%%%%%%%%%%%%%%%%%%%%%%%%%%%%%%%%%%%%%%%%%%%%%%%%%%%%%%%%%%%%%%% 

The skyrmion-vortex interaction is revealed by 
the dependence of the free energy of the system on the 
separation $r_{\rm sv}$ between the skyrmion and vortex cores.  
The free energy of an inhomogeneous superconductor is~\cite{Kosztin:prb98}:
\begin{equation}
F= -\frac{1}{\beta}\sum_n {\rm ln} \left[ 2 \cosh \left( \frac{ \beta \epsilon_n }{2} \right)  \right] + \frac{1}{V} \int {\rm d}\mathbf{r} |\Delta ( \mathbf{r} )|^2 ,  \label{Eq:F}
\end{equation}
where the sum is over states with positive energies $\epsilon_n$ and $\beta= 1/k_B T$. We focus on $T=0$~\cite{Comment:Temp} and calculate $F$ numerically for a  lattice of $41\times 29$ sites~\cite{Comment:Size}, with open boundary conditions at the edges. The ferromagnetic region covers a limited central region of $33 \times 21$ sites (dashed rectangle in Fig.~\ref{Fig3}b-c).
The chemical potential, exchange field, SOC and Debye frequency $\omega_D$~\cite{Comment:Debye} (scaled by the hopping energy $t$) are set to: $\mu/t= -4$, $h_0/t = -0.2$, $\alpha_R/t = 0.5$, and $\omega_D / t= 2.0$.

In Fig.~\ref{Fig3}a, we show the change of the free energy as a function of $r_{\rm sv}$ for a N\'{e}el skyrmion with $\varphi=0$.
For large SVP separations, $r_{\rm sv}\gtrsim 5R_{\rm s}$, the interaction is sufficiently weak that a small on-site pinning potential can be used to fix the location of the vortex relative to the skyrmion~\cite{Comment:Pinning}.
In this regime, we obtained an optimized vortex profile for each value of $r_{\rm sv}$ (Fig.~\ref{Fig3}a, red squares).
At smaller separations the skyrmion-vortex interaction is so strong that pinning becomes ineffective, and the vortex runs away to bind to the skyrmion.
Therefore, we estimate the interaction at short distances by first optimizing the vortex profile at $r_{\rm sv}=0$, and then use this fixed profile to calculate the change of free energy at larger separations (Fig.~\ref{Fig3}a, red circles).
In their region of overlap, the data from the two methods agree reasonably well, giving confidence in the approximation procedure.

The essential features of the SVP interaction may be summarized as follows:~(i) For separations smaller than the characteristic skyrmion radius $R_{\rm s}$, the free energy 
takes a harmonic form $ F\sim \frac{1}{2}kr_{\rm sv}^2$ where  $k\sim \kappa q_v h_0R_{\rm s}^{-1}\cos\varphi$. For the parameters chosen, we numerically find for the N\'{e}el skyrmion $k R_{\rm s}^2 /2t \sim 0.02$ (see Fig.~\ref{Fig3}a). (ii) For $r_{\rm sv}>R_{\rm s}$, the effective attractive force $-dF/dr$ softens substantially. 

In Figs.~\ref{Fig3}b,c, we verify the sign of the interaction by showing the ground state order parameter configuration for the case where the interaction is attractive ($\varphi=0$), and the skyrmion and vortex form a bound state (Fig.~\ref{Fig3}b), and where it is repulsive ($\varphi=\pi$), and the vortex is pushed to the edge of the ferromagnetic region (Fig.~\ref{Fig3}c). 
Note that since $|\Delta|$ is diminished by the exchange field, it is favorable for the vortex to remain in the magnetic region. 
We have also verified that the sign of the interaction indeed changes with the sign of the $\alpha_R$.

For the interaction between a vortex and a Bloch skyrmion, we have verified: (i) the sign of the interaction is {\it independent} of the sign of the $\alpha_R$ and its strength is systematically smaller than in the N\'{e}el case (both consistent with this being a second-order effect in $\alpha_R$),  and (ii) the interaction changes sign with $q_v h_0$, in agreement with Table~\ref{table1}. 
We provide the numerical analysis for the Bloch skyrmion in the Supplemental Material \cite{supp}.

We now investigate the dynamics of the composite N\'{e}el SVP. 
One particularly appealing approach for controlling the motion of this composite object  
is to utilize spintronics techniques. Below we determine the conditions under which the binding potential is sufficiently strong for the vortex to follow the skyrmion when it is driven by an external torque.
We identify a critical drift velocity above which the SVP dissociates.
  
We describe the motion of the composite SVP via semiclassical equations of motion for the skyrmion and vortex centers of mass.
An effective action capturing the motion of the skyrmion can be derived from the path integral formulation of the spin system~\cite{Makhfudz:prl12}; 
as is common practice we model the vortex dynamics using the conventional action of a massive particle subject to the Magnus force~\cite{Blatter94}.      
We are mainly interested in propagation for separations $r_{\rm sv} \lesssim R_{\rm s}$, 
and therefore model the skyrmion-vortex interaction by a harmonic potential $U_{\rm int}= \frac{1}{2}kr_{\rm sv}^2$ for $r_{\rm sv}<R_{\rm s}$ and $U_{\rm int}= \frac{1}{2}kR_{\rm s}^2$ for  $r_{\rm sv}>R_{\rm s}$.
Variation of the resulting total action and dissipation function yields the equations of motion (see Supplemental Information):
\begin{eqnarray}
m_s  \ddot{\boldsymbol{\mathcal{R}} }_s &=&-\mathbf{G}_s\times \left[ \dot{\boldsymbol{\mathcal{R}}}_s  - \mathbf{v}  \right] - 4\pi S \alpha_G \left[ \dot{\boldsymbol{\mathcal{R}}}_s  - \frac{\beta}{\alpha_G}\mathbf{v} \right]   \nonumber
\\ 
&&  - k\left[ \boldsymbol{\mathcal{R}}_s - \boldsymbol{\mathcal{R}}_v   \right],  \label{Eq:EOMs}\\
m_v  \ddot{\boldsymbol{\mathcal{R}} }_v &=& -\mathbf{G}_v\times \dot{\boldsymbol{\mathcal{R}}}_v  -\frac{\partial U_{\rm pin}}{\partial \boldsymbol{\mathcal{R}}_v  }  - \alpha_v \dot{\boldsymbol{\mathcal{R}} }_v \nonumber
\\ && + k \left[ \boldsymbol{\mathcal{R}}_s - \boldsymbol{\mathcal{R}}_v   \right]  . \label{Eq:EOMv} 
\end{eqnarray}
In Eqs.~(\ref{Eq:EOMs}-\ref{Eq:EOMv}) we assumed $r_{\rm sv}\equiv |\boldsymbol{\mathcal{R}}_s - \boldsymbol{\mathcal{R}}_v|<R_{\rm s}$, where $\boldsymbol{\mathcal{R}}_s$ ($\boldsymbol{\mathcal{R}}_v$) is the center of mass position of the skyrmion (vortex) and $m_s$ ($m_v$) its mass.
The prefactors are $\mathbf{G}_s= 4\pi S q_s \hat{\mathbf{z}}$ and $\mathbf{G}_v= 2\pi n_s q_v \hat{\mathbf{z}}$, where $S$ and $n_s$ are the spin density and the superfluid density, respectively.    
$U_{\rm pin}$ represents a vortex pinning potential, while the vector $\mathbf{v}$ 
arises from adiabatic torques due to electric currents or thermal gradients. The dissipative processes are parametrized by  the Gilbert damping parameter $\alpha_G$ and the friction constant $\alpha_v$ of the vortex, whereas $\beta$ determines the non-adiabatic torque.

The dynamics of Eqs.~(\ref{Eq:EOMs}-\ref{Eq:EOMv}) can be readily characterized 
when the vortex motion is overdamped, and the SVP rapidly enters a steady state regime, driven by the external torque $\mathrm{\bf{v}}$. Setting $\ddot{\boldsymbol{\mathcal{R}} }_s= \ddot{\boldsymbol{\mathcal{R}} }_v=0$ and $\dot{\boldsymbol{\mathcal{R}} }_s=\dot{\boldsymbol{\mathcal{R}} }_v= \dot{\boldsymbol{\mathcal{R}}}$ in Eqs.~(\ref{Eq:EOMs}-\ref{Eq:EOMv}), we find (setting $U_{\mathrm{pin}}=0$ for now) 
\begin{eqnarray}
\label{Eq:velocity}
|\dot{\boldsymbol{\mathcal{R}}}|=|\mathbf{v}|\sqrt{\frac{q_s^2+\beta^2}{(q_s+q_v n_s/2 S)^2+(\alpha_G+\alpha_v/4\pi S)^2}}.
\end{eqnarray}

The angle $\gamma$ between the SVP velocity and the external torque, $\dot{\boldsymbol{\mathcal{R}}}\cdot\mathrm{\bf{v}}\propto\cos\gamma$, may be expressed as 
\begin{eqnarray}
\label{Eq:angle}
\mathrm{tan}\,\gamma=\frac{q_s(\alpha_G-\beta+\alpha_v/4\pi S)-n_s q_v \beta/2 S}{q_s^2+(\alpha_G+\alpha_v/4\pi S)\beta+n_s  q_v q_s/2 S}.
\end{eqnarray}
We generally find $\gamma\neq0$ due to the effective Lorentz and Magnus forces induced on the skyrmion and vortex. 

Due to the softening of the binding potential at large distances $r_{\rm sv} \gtrsim R_{\rm s}$, see Fig.~\ref{Fig3}a, the SVP dynamically unbinds when the steady-state separation $\bar{r}_{\rm sv}$ exceeds $R_{\rm s}$.
This condition yields a critical value for $ |\mathbf{v} |$
\begin{equation}
v^+\sim \frac{kR_{\rm s}}{2\pi}\sqrt{\frac{(q_s+q_v n_s/ 2 S)^2+(\alpha_G+\alpha_v/4\pi S)^2}{(n_s^2q_v^2+(\alpha_v/2\pi)^2)(q_s^2+\beta^2)}},
\end{equation}
where the SVP dissociates.

The presence of a vortex pinning potential, characterized by a finite spring constant $k_{\rm pin}$ within the pinning radius $l_{\rm pin}$, may also trap the bound state if the applied torque is too weak, $|\mathrm{\bf{v}}|<v^-$. The depinning torque $v^-$ can be estimated by determining when the static solution $\ddot{\boldsymbol{\mathcal{R}}}_{\rm s/v}=\dot{\boldsymbol{\mathcal{R}}}_{\rm s/v}=0$ to Eqs.~(\ref{Eq:EOMs}-\ref{Eq:EOMv}) becomes unstable:
\begin{equation}
v^-\sim\frac{k_{\rm pin}l_{\rm pin}}{4\pi S\sqrt{q_s^2+\beta^2}}.
\end{equation}
Thus, we find that within a characteristic window of applied torque, $v^-<|\mathbf{\bf{v}}|<v^+$, the 
SVP is depinned and propagates as a bound pair described by Eqs.~(\ref{Eq:velocity}-\ref{Eq:angle}).
This window closes when the pinning force exceeds the characteristic value $k_{\rm pin}^*l_{\rm pin}^*\sim 2 S kR_{\rm s}\sqrt{\frac{(q_s+q_v n_s/2 S)^2+(\alpha_G+\alpha_v/4\pi S)^2}{n_s^2q_v^2+(\alpha_v/2\pi)^2}}$, in which case the vortex remains pinned for any applied torque. 
  
%%%%%%Summary%%%%%%%%%%%%%%%%%%%%
\emph{Discussion} $\--$
We expect SVPs to form in heterostructures based on materials with strong SOC. 
For example, transition metal dichalcogenides (TMDs) provide an interesting starting point for building such heterostructures due to their stackable layered structures, strong intrinsic SOC, and recently observed superconductivity ($T_c \sim 3$~K) even down to the atomically-thin limit~\cite{Xi:np16}. Importantly, the vortex phase of these systems can coexist with the skyrmion phase of ultrathin magnetic films in a compatible range of temperatures and magnetic field strengths. The TMDs are type-II superconductors, in which the formation of vortices is observed for applied magnetic fields 0.1 T $< B <$ 0.75 T, and are expected to exist up to 4.0 T~\cite{Xu:prl14, Zehetmayer:sr15}. 
As a promising exemplary candidate for the magnetic layer, a recent experiment on PdFe bilayers on Ir(111) showed that single N\'{e}el skyrmions can been written and deleted using spin-polarized currents, 
in an external magnetic field of $B=1.8$~T~\cite{Romming:science2013}. The skyrmions exist at temperatures from $0$~K~\cite{Simon:prb14} to above $T_c$~\cite{Romming:science2013} and have a size of $R_{\rm s}\sim 7$~nm. Thus, a heterostructure consisting of TMD and PdFe/Ir layers presents a promising platform studying the formation of SVPs. In the Supplemental Material, we provide an analytic estimate of the SVP binding energy in such a system and find that exchange fields as small as $h_0\sim 2.3$~meV give a binding energy between $\sim 0.3-3$~K, depending on the value of the Rashba SOC. The collective motion of the SVPs can for example be detected via imaging using high-resolution nano-magnetometry~\cite{Zeldov:NanoSQUID, Thiel:arxiv15}.  

In cases where the superconductor is in a topologically non-trivial phase, the SVP may bind a Majorana zero mode (cf.~\cite{Bjornson:prb2013, Pershoguba2015, Loss2016}).
In this case, spintronic techniques for manipulating the motion of skyrmions may be used to move the Majorana-carrying SVPs in a controlled way.
Identifying specific materials and high-precision control techniques for realizing and manipulating SVPs present interesting directions for future work.

\acknowledgements{{\it Acknowledgements} $\--$
We gratefully acknowledge the support of the Danish National Research Council, the Danish Council for Independent Research | Natural Sciences, the Villum Kann Rasmussen Foundation, and the People Programme (Marie Curie Actions) of the European Union’s Seventh Framework Programme (FP7/2007-2013) under REA Grant Agreement No. PIIF-GA-2013-627838.}
%%% References %%%%%%%%%%%%%%%%%%%%%%%%%%%%%%%%%%%%%%%%%%%%%%%%%%%%%%%%%%%% 

\begin{widetext}
\section{Supplemental Material}
\subsection{Equations of motion for coupled skyrmion-vortex dynamics}
The total action of the coupled skyrmion-vortex system can be written as
\begin{equation}
S= S_s + S_v + S_I,  \label{Eq:Stot}
\end{equation}
where $S_s$ and $S_v$ denote the action of the isolated skyrmion and vortex, respectively, whereas $S_I$ describes the coupling.

An effective action for the center of mass position $\boldsymbol{\mathcal{R}}_s= (r_x^s, r_y^s)$ of the skyrmion can be derived by substituting a skyrmion ansatz into the path integral formulation of the spin system and integrating out fluctuations and the spatial coordinate~\cite{Makhfudz:prl12}:
\begin{equation}
S_s= \int {\rm dt} \left[ \frac{1}{2}m_s \dot{\boldsymbol{\mathcal{R}}_s}^2  + \boldsymbol{\mathcal{A}}_s (\boldsymbol{\mathcal{R}}_s) \cdot \left( \dot{\boldsymbol{\mathcal{R}}_s}  - \mathbf{v} \right)   \right] . \label{Eq:S_s}
\end{equation}
Here, $m_s$ is the skyrmion mass,  $\boldsymbol{\mathcal{A}}_s  (\boldsymbol{\mathcal{R}}_s) $ the Berry-phase gauge potential, which satisfies $\boldsymbol{\nabla}_{\boldsymbol{\mathcal{R}}_s}\times \boldsymbol{\mathcal{A}}_s  = 4\pi S q_s \hat{\mathbf{z}}\equiv \mathbf{G}_s $, with $S$ the spin density of the ferromagnet in the two-dimensional $xy$-plane, and $\mathbf{v}$ arises from the induced torque. 
In metallic systems, $\mathbf{v}$ is proportional to the applied current density~\cite{Ralph:jmmm08}.  
For ferromagnetic insulators, in which skyrmions are driven by spin waves and thermally-induced torques~\cite{Bauer:nm12}, there are two torque contributions:
one term where $\mathbf{v}$ is proportional to the magnon current density ($\mathbf{j}_m$), i.e., $\mathbf{v}\propto \mathbf{j}_m$,  and a second term (due to Brownian motion of the skyrmion) that is proportional to the temperature gradient, i.e., $\mathbf{v}\propto \boldsymbol{\nabla} T$ ~\cite{Kovalev:epl12, Kong:prl13}.
 
The action of an isolated vortex is \cite{Blatter94, Nikolic06}
\begin{equation}
S_v= \int {\rm dt} \left[ \frac{1}{2}m_v \dot{\boldsymbol{\mathcal{R}}_v}^2  + \boldsymbol{\mathcal{A}}_v (\boldsymbol{\mathcal{R}}_v) \cdot  \dot{\boldsymbol{\mathcal{R}}_v}   - U_{\rm pin} (\boldsymbol{\mathcal{R}}_v)  \right] , \label{Eq:S_v}
\end{equation}
where $\boldsymbol{\mathcal{R}}_v= (r_x^v, r_y^v )$ is its center of mass position, $m_v$ is the mass of the vortex, $\boldsymbol{\nabla}_{\boldsymbol{\mathcal{R}}_v}\times \boldsymbol{\mathcal{A}}_v  = 2\pi\hbar n_s q_v \hat{\mathbf{z}}\equiv \mathbf{G}_v $, with $n_s$ the density of Cooper pairs, and $U_{\rm pin} (\boldsymbol{\mathcal{R}}_v) $ represents a pinning potential.
We have in Eq.~\eqref{Eq:S_v} disregarded the elastic energy associated with deformations of the straight vortex line, which is small for a thin film.  

For small separations of the skyrmion and vortex, the coupling term can be written as
\begin{equation}
S_I= -\frac{k}{2}\int {\rm dt}  \left[ \boldsymbol{\mathcal{R}}_s - \boldsymbol{\mathcal{R}}_v   \right]^2 . \label{Eq:S_I}
\end{equation}

The dissipation of the skyrmion dynamics is found by substituting a skyrmion ansatz into the dissipation functional of the magnetic system~\cite{Gilbert04}, followed by a spatial integration: 
\begin{equation}
\Gamma_s= 2\pi S \alpha_G \int {\rm dt}  \left[ \dot{\boldsymbol{\mathcal{R}} }_s - \frac{\beta}{\alpha_G}\mathbf{v} \right]^2 . \label{Eq:R1}
\end{equation}
Here, $\alpha_G$ is the Gilbert damping parameter and $\beta$ is the non-adiabatic torque parameter. As mentioned above, there are two torque contributions when the skyrmion is driven by thermal torques, i.e., two different $\beta$-terms.   A standard Rayleigh dissipation function captures the friction processes of the vortex dynamics 
\begin{equation}
\Gamma_v= \frac{\alpha_v}{2}\int {\rm dt}\,  \dot{\boldsymbol{\mathcal{R}} }_v^2  , \label{Eq:R2}
\end{equation}
where $\alpha_v$ is the friction parameter. The total dissipation function is $\Gamma= \Gamma_s + \Gamma_v$.

The equations of motion for the vortex and the skyrmion are found from the variational equations $\delta S/ \delta \boldsymbol{\mathcal{R}}_i = \delta \Gamma/ \delta \dot{\boldsymbol{\mathcal{R}}}_i $  ($i\in \{ s, v \}$):
\begin{eqnarray}
m_s  \ddot{\boldsymbol{\mathcal{R}} }_s &=&-\mathbf{G}_s\times \left[ \dot{\boldsymbol{\mathcal{R}}}_s  - \mathbf{v}  \right] - 4\pi S \alpha_G  \left[ \dot{\boldsymbol{\mathcal{R}}}_s  - \frac{\beta}{\alpha_G}\mathbf{v} \right]  - 
k \left[ \boldsymbol{\mathcal{R}}_s - \boldsymbol{\mathcal{R}}_v   \right]  ,  \label{Eq:EOMs2}  \\
m_v  \ddot{\boldsymbol{\mathcal{R}} }_v &=& -\mathbf{G}_v\times  \dot{\boldsymbol{\mathcal{R}}}_v -\frac{\partial U_{\rm pin}}{\partial \boldsymbol{\mathcal{R}}_v  }  - \alpha_v \dot{\boldsymbol{\mathcal{R}} }_v + k \left[ \boldsymbol{\mathcal{R}}_s - \boldsymbol{\mathcal{R}}_v   \right]   .  \label{Eq:EOMv2} 
\end{eqnarray}
Eqs.~\eqref{Eq:EOMs2}-\eqref{Eq:EOMv2} yield an effective description of the coupled skyrmion-vortex dynamics.

\subsection{Interaction between a Bloch skyrmion and a vortex }
Fig.~\ref{Fig1S} shows the change of the free energy for different separations between a vortex and a Bloch skyrmion  (cf.~main text for the case of a N\'{e}el skyrmion).
Note that the values based on a fixed vortex profile largely overestimate the change of the free energy compared to the values calculated at large distances where the vortex is pinned and its profile solved for self-consistently.
As we explain below, this is due to a strong dependency of the vortex profile on the separation between the Bloch skyrmion and the vortex. 

\begin{figure}[ht] 
\centering 
\includegraphics[scale=1.2]{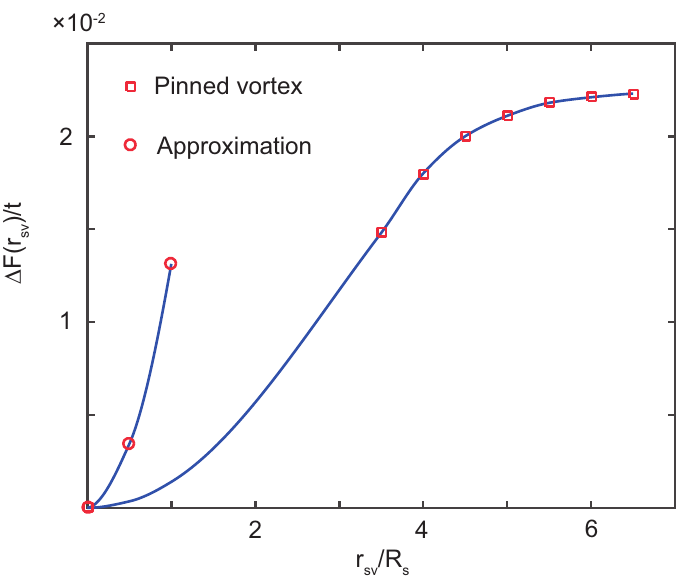}  
\caption{(Color online).  The change of the free energy $\Delta F (r_{\rm sv})= F(r_{\rm sv}) - F_{\rm eqv}$ as a function of the separation $r_{\rm sv}$ between a vortex and a Bloch skyrmion. $F_{\rm eqv}$ is the equilibrium value of the free energy at $r_{\rm sv}=0$.  The circles represent values calculated by assuming a fixed vortex profile that is calculated self-consistently at $r_{\rm sv}=0$. The squares represent a pinned vortex whose shape is determined self-consistently for each separation, while the lines are guides to the eye.
The material parameters are specified in the main text.}
\label{Fig1S} 
\end{figure} 
 
Figure \ref{Fig2S}a shows the normalized phase vector $({\rm Re} [ \Delta (\mathbf{r}) ], {\rm Im} [ \Delta (\mathbf{r}) ] )$ of the pair potential $\Delta (\mathbf{r}) $ around a bound state formed by a vortex and a Bloch skyrmion.
Because of the in-plane structure of the Bloch skyrmion, the pair potential develops a non-trivial phase dependence along the radial direction away from the vortex core. 

\begin{figure}[ht] 
\centering 
\includegraphics[scale=0.4]{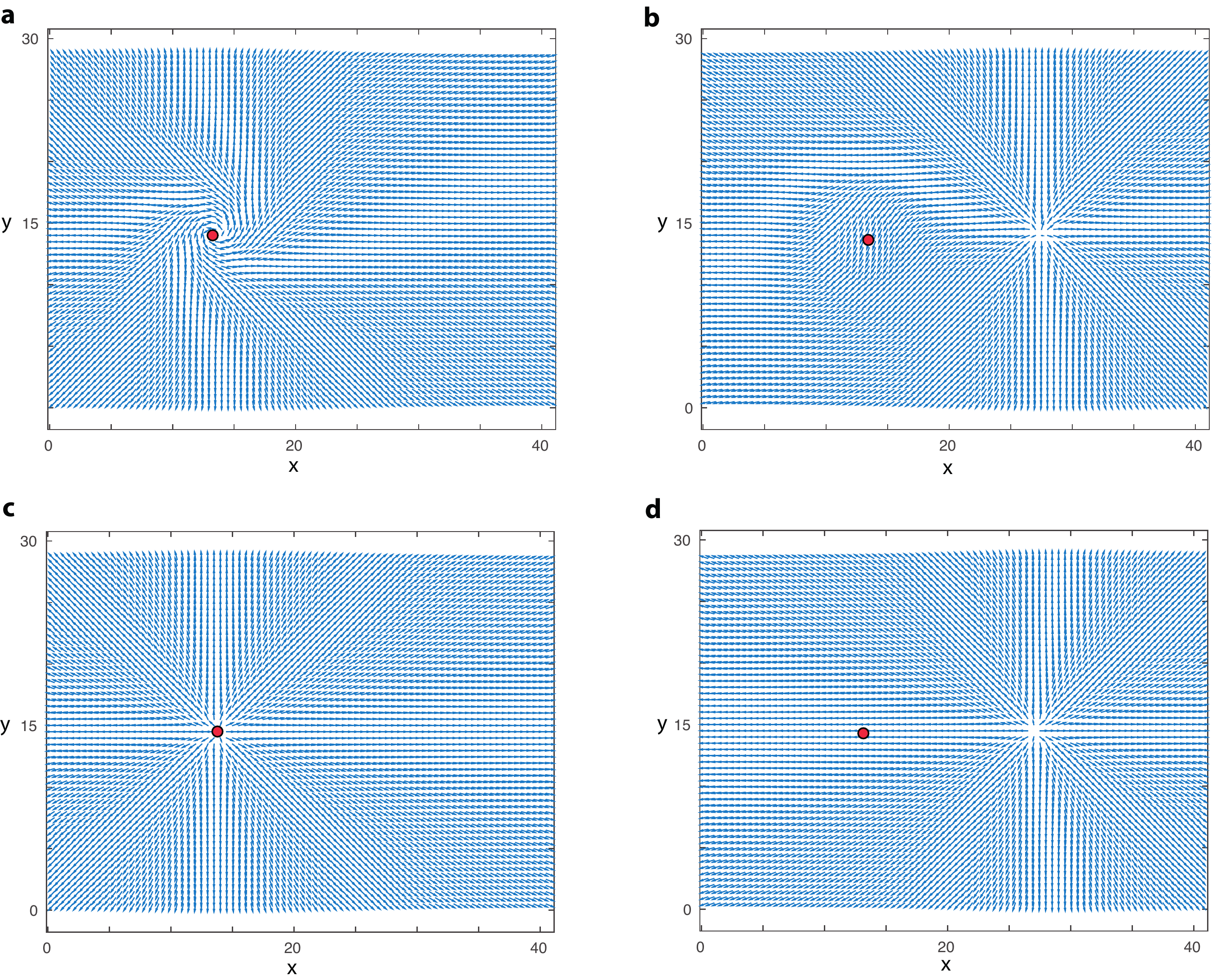}  
\caption{(Color online). {\bf (a)} Self-consistent solution of a vortex bound to a Bloch skyrmion. {\bf (b)} Self-consistent solution of a vortex that is pinned far away from the Bloch skyrmion.
 {\bf (c)} Self-consistent solution of a vortex that is bound to a N\'{e}el skyrmion. {\bf (d)} Self-consistent solution of a vortex that is pinned far away from the N\'{e}el skyrmion.
 In all figures, the red dot indicates the location of the skyrmion center and the phase of the pair potential is represented by the vector
 $({\rm Re} [ \Delta (\mathbf{r}) ], {\rm Im} [ \Delta (\mathbf{r}) ] ) / \sqrt{  {\rm Re} [ \Delta (\mathbf{r}) ] ^2 +  {\rm Im} [ \Delta (\mathbf{r}) ]^2  }$. }
\label{Fig2S} 
\end{figure} 

For a Bloch skyrmion, the in-plane component of the magnetization is perpendicular to the radial direction, i.e.,
$h_x(r,\phi) \sim -\sin (\phi)$ and $h_y(r,\phi) \sim \cos (\phi)$, which for an unmodified vortex phase profile gives rise to an anomalous supercurrent density ${\bf j}_{\rm me}\sim \kappa \left(\hat{z}\times {\bf h}\right)$ in the radial direction. 
Consequently, in order to produce a total supercurrent density with no divergence (as required by the continuity equation), the vortex pair potential must develop a radial phase dependence. 
For a vortex that is pinned far away from the Bloch skyrmion, the phase around the vortex attains a conventional form with only small variations along the radial direction (Fig.~\ref{Fig2S}b).
These results demonstrate that self-consistent solution of the pairing potential is essential for capturing the physics of Bloch skyrmion-vortex binding.

In contrast to the case above, N\'{e}el skyrmions readily produce a divergenceless anomalous supercurrent around the vortex-skyrmion bound state and no modulations of the phase along the radial direction appear (Fig.~\ref{Fig2S}c). 
The vortex will therefore maintain a fixed profile for different separations $r_{\rm sv}$ (Fig.~\ref{Fig2S}c-d), and an approximate solution considering a fixed phase profile for varying $r_{\rm sv}$ may yield reasonable results (as seen in the main text).

\subsection{Transient dynamics of a Skyrmion-Vortex bound pair}
In Fig.~\ref{Fig3S} we plot a typical skyrmion-vortex pair trajectory, obtained by solving the coupled equations of motion, Eqs.~(6-7) of the main text.  The individual velocities (positions) of the skyrmion and the vortex are shown in the main panel (inset). % of the skyrmion and vortex velocities (positions, inset), parametrized by the evolution time $t$, 
%that result from solving their dynamical equations of motion, Eqs.~(6-7) of the main text. 
After a short transient time, the bound pair reaches a steady state with a velocity given by Eqs.~(8-9) in the main text (indicated by the dashed vector in Fig.~\ref{Fig3S}). The inset of Fig.~\ref{Fig3S} shows the time-evolved skyrmion and vortex positions. The two circles indicate the skyrmion and vortex positions at a given time, with a separation magnitude $r_{\rm sv}=0.21 R_{\rm s}$ that is close to the steady state value for the parameter set used (see Fig.~\ref{Fig3S} caption). The relative angle $\gamma^\prime$ between the separation vector and the velocity results from the vortex Magnus force in Eq.~(7) of the main text. The steady state solution is given by $\mathrm{tan}\gamma^\prime=2\pi n_sq_v/\alpha_v$, as indicated in the inset. % and is shown in the inset of Fig.~\ref{Fig3}. 
We have checked that other parameter sets (e.g. $m_s/m_v\sim1$) lead to qualitatively similar transient dynamics, while the long-time steady states converge to the predictions of Eqs.~(8-9) of the main text for $|{\bf v}|<v_+$.

\begin{figure}[h!] 
\centering 
\includegraphics[scale=0.35]{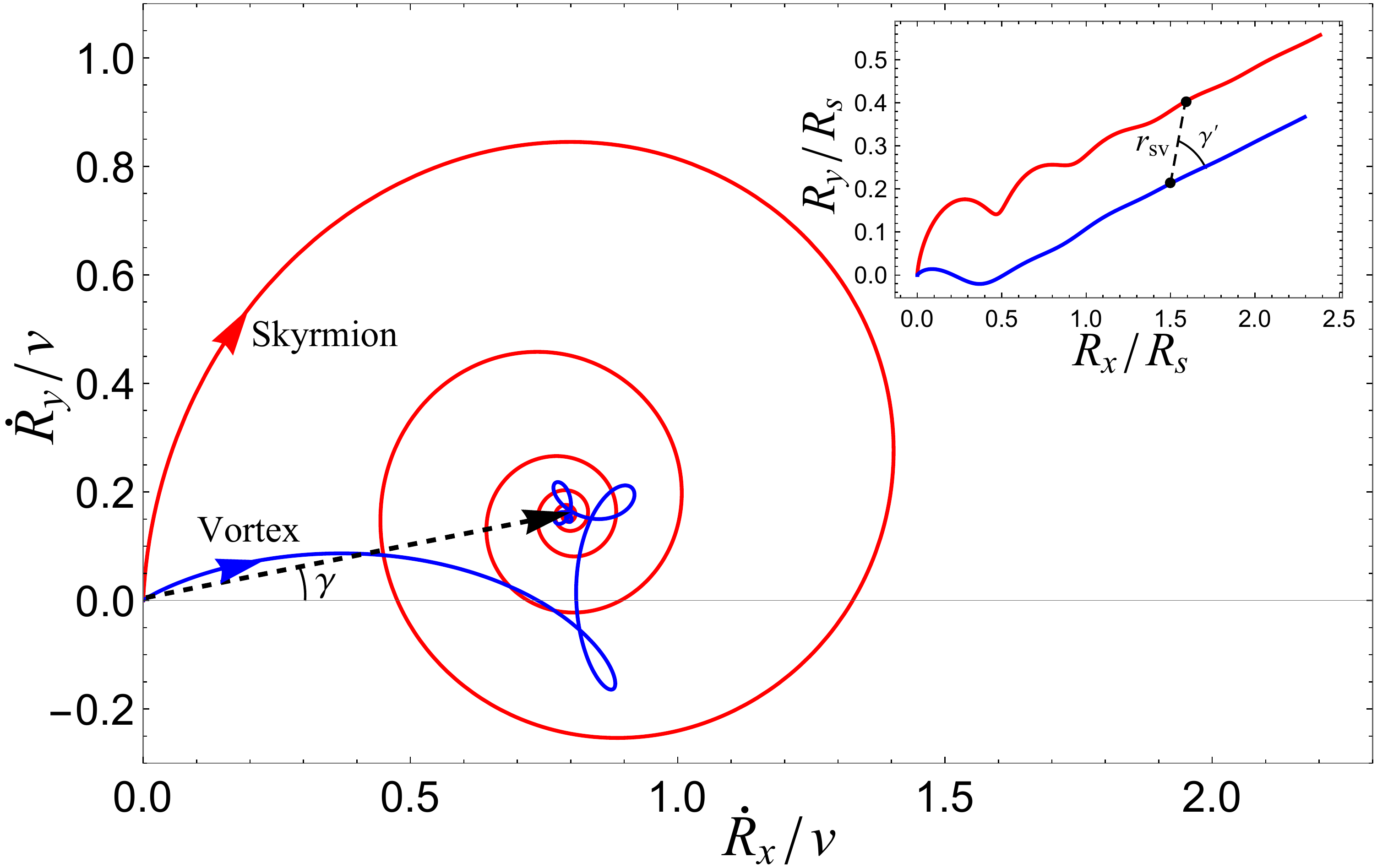}  
\caption{(Color online). Skyrmion and vortex trajectories (red and blue, respectively) determined by solving Eqs.~(6-7) of the main text in the absence of the pinning potential, $U_{\rm pin}=0$. We initialized both the skyrmion and the vortex at the origin with zero velocity and used the parameters $|{\bf\rm G_s}|/|{\bf\rm G_v}|=5,\,|{\bf\rm G_v}|/\alpha_v=1.33,\,|{\bf\rm G_s}|/4\pi S\alpha_s=6.66,\,4\pi S\alpha_s/\alpha_v=1,\,\alpha_s/\beta=3,\,m_s/m_v=10^3,\,kR_s^2/m_s |{\bf\rm v}|^2=10$.}
\label{Fig3S} 
\end{figure} 

\subsection{Numerical solution of the Bogoliubov-de Gennes equations}
We model the system by the tight binding Hamiltonian
\begin{eqnarray}
H &=& -t \sum_{\langle ij \rangle} \mathbf{c}_{i}^{\dagger} \mathbf{c}_{j} - \mu\sum_{i} \mathbf{c}_{i}^{\dagger} \mathbf{c}_{i} - \sum_{i}  \mathbf{c}_{i}^{\dagger} \left(  \mathbf{h}_i \cdot\boldsymbol{\sigma} \right) \mathbf{c}_{i} +  i\alpha_R \sum_{\langle ij \rangle}  \mathbf{c}_{i}^{\dagger}\hat{\mathbf{z}}\cdot (  \hat{\mathbf{d}}_{ij} \times \boldsymbol{\sigma} ) \mathbf{c}_{j} + \sum_{i} \left( \Delta_i  c_{i\uparrow}^{\dagger} c_{i\downarrow}^{\dagger} +  h.c. \right)  ,    \label{Eq:H0S} 
\end{eqnarray}
where the chemical potential, exchange field,  and SOC  (scaled by the hopping energy $t$) are set to: $\mu/t= -4$, $h_0/t = -0.2$,  and $\alpha_R/t = 0.5$, respectively.
The Hamiltonian \eqref{Eq:H0S} can be related to the corresponding continuum model by using the central difference approximation.  In this approximation, the energies $t$ and $\alpha_R$  are given by
$ t= \hbar^2/2ma^2$ and $\alpha_{R}/ t = ma\tilde{\alpha}_{R}/\hbar^2$, where $a$ is the spacing between the lattice points, $m$ is the effective mass, and $\tilde{\alpha}_{R}$ is the SOC parameter of the continuum model.   
The parameter values given above model a lightly hole-doped semiconductor, in which 
the effective mass is $m=0.6 m_e$ ($m_e$ is the electron mass), the SOC is $\tilde{\alpha}_{R}= 0.21$~eV\AA, the exchange field is $h_0=1.4$~meV, and the Fermi energy is $E_F= 3.17$~meV when measured from the bottom of the lowest subband~\cite{Stormer:prl83}. The discretization constant of this system is set to $a= 3$~nm, which is much smaller than the Fermi wavelength $\lambda_F \sim 19$ nm.
This ensures a stable calculation of the eigenvectors and eigenvalues of the Bogoliubov-de Gennes Hamiltonian.

We solve the pair potential $\Delta_{\mathbf{i}}= V \langle c_{\mathbf{i} \uparrow} c_{\mathbf{i} \downarrow}  \rangle$ of the superconductor self-consistently.
By inserting the Bogoliubov transformation  $c_{\mathbf{i}\tau}= \sum_n [  u_{n\tau} (\mathbf{i}) \gamma_n +  v_{n\tau}^{\ast} (\mathbf{i}) \gamma_n^{\dagger} ]$  into the expression for the pair potential and taking the thermal average, we find the following self-consistency condition:
\begin{eqnarray}
\Delta_{\mathbf{i}} &=& -\frac{V}{2}\sum_{n\tau\tau^{'}} (i\sigma_y)_{\tau \tau^{'}} v_{n\tau}^{\ast} (\mathbf{i}) u_{n\tau^{'}} (\mathbf{i}) \left[ 1-2f (\epsilon_n) \right] .\label{Eq:delta1}
\end{eqnarray}
Here, $V$ is the pairing energy, and $\gamma_n^{\dagger}$ ($\gamma_n$) are the Bogoliubov quasi-particle creation (destruction) operators, which represent a complete set of energy eigenstates,  $H= E_g + \sum_n \epsilon_n \gamma_n^{\dagger}\gamma_n $ ($E_g$ is the groundstate energy), and $f(\epsilon)$ is the Fermi-Dirac distribution. The summation runs over positive energy eigenstates with an energy smaller than the cut-off energy $\hbar\omega_D= 2t$  set by the Debye frequency $\omega_D$~\cite{Sacramento:prb07}. For the simulation, we take $V = 5t$.
The pair potential is iteratively solved together with the self-consistency condition \eqref{Eq:delta1} until the Euclidean norm of the pair potential ($\| \Delta \| = \sqrt{\sum_{\mathbf{i}} |\Delta_{\mathbf{i}}  |^2}$) reaches a relative error on the order $10^{-5}$. \\

\subsection{Analytic estimate of binding energy}
The magnetoelectric coupling parameter $\kappa$ is related to the SOC in the microscopic Hamiltonian via the relationship~\cite{balatsky} 
\begin{equation}
\kappa = \frac{m \tilde{\alpha}_R}{ 2 \pi \hbar^2 }.
\end{equation} 
From Eq. (3) in the Letter, we find that the binding energy of a SVP is $F_{\rm me} (0) = 2 \pi \kappa h_0 R_s$. 

We now estimate the binding energy for an example pair of superconducting and magnetic materials  that coexist in their vortex and skyrmion supporting phases within the same range of temperatures and magnetic fields.
In NbSe$_2$,  an atomically-thin intrinsic superconductor, the total bandwidth is $8t\approx 1$~eV and  the lattice constant is $a= 3.5$~\AA~\cite{Johannes:prb06}. In the valence band of a TXY (where T stands for a transition metal atom, and X and Y stand for chalcogen atoms) heteromonolayer, $\tilde{\alpha}_R$ typically takes values in the range of $2-14$~meV\AA~\cite{Cheng:epl13,Kuc2015}. For a heterobilayer (e.g., ${\rm MoS_2}$ on ${\rm W Se_2}$), the Rashba coupling $\tilde{\alpha}_R\sim 0.2-1.4$~meV\AA\, is typically smaller. For ${\rm NbSe}_2$ on a substrate, we thus expect $\tilde{\alpha}_R\sim 0.5$~meV\AA\,, or even larger $\tilde{\alpha}_R\sim$~5 meV\AA\, for a heteromonolayer NbXY. We also note that applying an out-of-plane electric field of order $0.1-0.2$~eV/\AA\, could allow one to tune $\tilde{\alpha}_R$ substantially \cite{Kuc2015}, perhaps even reversing its sign.
Using the conservative estimate $\tilde{\alpha}_R=0.5$~meV\AA\, for ${\rm NbSe_2}$ on a substrate, and assuming $R_s=7$ nm with an effective mass $m$ related to $t$ using $ t= \hbar^2/2ma^2$, we find that $F_{\rm me} (0)/k_B = 300$ mK when $h_0= 2.3$ meV.

\end{widetext}


\begin{thebibliography}{99}
\bibitem{Comment:Dresselhaus} However, if there is also Dresselhaus SOC, a SVP with $q_{\rm s}=-1$ may occur since the magnetoelectric current in this case winds around the skyrmion core, ${\bf j}_{\rm Dress}\propto S_x\hat y+S_y\hat x$, similar to the case $q_{\rm s}=1$ in the presence of Rashba SOC.
\bibitem{Comment:h}Throughout we assume that the exchange interaction within the ferromagnet is sufficiently large so as to neglect determining $\mathbf{h}$ self-consistently in the presence of the vortex.
\bibitem{Comment:Temp}To facilitate rapid convergence, we have used a small temperature $T=0.001 t$ which is far below the  quasiparticle gap. We have checked that our results are not sensitive to this choice.
\bibitem{Comment:Size} The size of the system is chosen such that the centers of the vortex and the skyrmion (in all calculations) are located at a distance larger than $6R_{\rm s}$ from the sample edges. This ensures that the calculation of the vortex profile and the free energy change $\Delta F $  are not influenced by the edges (i.e., no finite size effects), as we have confirmed explicitly by varying the system size.
\bibitem{Comment:Pinning} We have checked that changing the strength of the pinning potential has a negligible effect on $\Delta F (r_{\rm sv})$.
\end{thebibliography}
\end{document}